\documentclass[preprint]{iucr}
     \journalcode{A}
\usepackage{enumitem}
\usepackage{xspace}
\usepackage{fancyhdr}
\usepackage{afterpage}
\usepackage[usenames,dvipsnames]{xcolor}
\usepackage{dcolumn}
\usepackage{comment}
\usepackage{listings}
\usepackage{makeidx}
\usepackage{longtable}
\usepackage{chemformula}
\usepackage[utf8]{inputenc}
\usepackage{comment}

\usepackage[dvips]{epsfig}
\usepackage{epstopdf}
\usepackage{graphicx}

\usepackage{booktabs}

\usepackage{mathtools}
\usepackage{amsmath}
\usepackage{amssymb}
\usepackage[version=4]{mhchem}
\usepackage{bm}
\usepackage{amsfonts}
\usepackage{commath} 
\usepackage{gensymb}
\usepackage[a4paper, total={6in, 9in}]{geometry}
\usepackage{adjustbox}

\usepackage{subfiles}

\usepackage[frozencache=true]{minted}

\setminted{
  frame=single,
  framesep=5pt,
  fontsize=\small,
  breaklines=true
}

\usepackage{hyperref}

\usepackage[most]{tcolorbox}
\usepackage{xcolor}
\tcbset{colframe=white, coltitle=black, boxrule=0pt, sharp corners}

\newcommand{\sjba}[1]{}

\newcommand{\mta}[1]{}

\newcommand{\fig}[1]{Fig.~\ref{fig:#1}}

\newcommand{\sect}[1]{Section~\ref{sec:#1}}

\newcommand{\tabl}[1]{Table~\ref{tab:#1}}

\graphicspath{ {./figures/} }

\definecolor{dgreen}{HTML}{008000}

\definecolor{codegreen}{rgb}{0,0.6,0}
\definecolor{codegray}{rgb}{0.5,0.5,0.5}
\definecolor{codepurple}{rgb}{0.58,0,0.82}
\definecolor{backcolour}{rgb}{0.95,0.95,0.92}
\lstdefinestyle{mystyle}{
    backgroundcolor=\color{backcolour},
    commentstyle=\color{codegreen},
    keywordstyle=\color{magenta},
    numberstyle=\tiny\color{codegray},
    stringstyle=\color{codepurple},
    basicstyle=\ttfamily\footnotesize,
    breakatwhitespace=false,
    breaklines=true,
    captionpos=b,
    keepspaces=true,
    numbers=left,
    numbersep=5pt,
    showspaces=false,
    showstringspaces=false,
    showtabs=false,
    tabsize=2
}
% ...
\newcounter{saveenumi}

\newcommand{\be}{\begin{enumerate}}
\newcommand{\ee}{\end{enumerate}}
\newcommand{\bes}{\begin{enumerate}[wide, labelwidth=!, labelindent=0pt, label=\textbf{\textcolor{blue}{\arabic*}.}]}
\newcommand{\ees}{\end{enumerate}}

\newcommand{\utils}{{\sc diffpy.utils}\xspace}

\newcommand{\structure}{{\sc diffpy.structure}\xspace}

\newcommand{\skpkg}{{\sc scikit-package}\xspace}
\newcommand{\function}{{\tt function}\xspace}
\newcommand{\module}{{\tt module}\xspace}
\newcommand{\workspace}{{\tt workspace}\xspace}
\newcommand{\system}{{\tt system}\xspace}
\newcommand{\public}{{\tt public}\xspace}

\newcommand{\skbuild}{{\sc scikit-build}\xspace}

\newcommand{\precommit}{{\sc pre-commit}\xspace}

\newcommand{\level}[1]{\hyperref[sec:level#1]{Level~#1}}

\begin{document}

\title{scikit-package - software packaging standards and
roadmap for sharing reproducible scientific software}
\author[a]{Sangjoon}{Lee}{}{}
\author[a]{Caden}{Myers}{}{}
\author[a]{Andrew}{Yang}{}{}
\author[a]{Tieqiong}{Zhang}{}{}
\cauthor[a]{Simon~J.~L.}{Billinge}{sb2896@columbia.edu}{}

\aff[a]{Department of Applied Physics and Applied Mathematics, Columbia University, \city{New York}, NY~10025, \country{USA}}
\shortauthor{}

\maketitle

\begin{abstract}
Scientific advancement relies on the ability to share and reproduce results.
When data analysis or calculations are carried out using software written by scientists there are special challenges around code versions, quality and code sharing.
\skpkg provides a roadmap to facilitate code reuse and sharing with minimal effort through tutorials coupled with automated and centralized reusable workflows. 
The goal of the project is to provide pedagogical and practical tools for scientists
who are not professionally trained software engineers to write more reusable and maintainable software code.
Code reuse can occur at multiple levels of complexity—from turning a code block into a function within a single script, to publishing a publicly installable, fully tested, and documented software package 
\skpkg provides a community maintained set of tools, and a roadmap, to help scientists bring their software higher levels of reproducibility and shareability.
\end{abstract}


\section{Introduction}

Software code is widespread in modern science. 
A challenge for scientists is reusing and sharing code, something that is fundamental for scientific reproducibility.
If you are a scientist with code to share, where should you begin?
Sharing code can be challenging due to the effort required to learn the software community's best practices. 
\skpkg is designed to lower this barrier and provide tools for increasing the uniformity and quality of shared code.

Sharing can be done across levels of complexity, from simple reuse of functions within a file, all the way to releasing community installable software packages.
Here, we present a roadmap and helper software at each levels to lower the barrier for scientists to adopt these practices.
Since most scientific software is in the Python language, we build tools around this, though the general principles and tools can be applied more broadly.

There are multiple resources for packaging Python software, but in our view these are are less easy to follow for non-specialists.  
The Python Packaging Authority (PyPA) is a working group that drafts and hosts online documentation~\cite{authoritypypaPythonPackagingUser} on how to package, share, and install Python software. However, mastering this infrastructure from scratch consumes considerable time that could otherwise be devoted to writing scientific code. There have been various attempts to build tools for package management, such as {\sc Poetry}~\cite{eustacePoetryPythonPackaging2025} that facilitates the creation of a boilerplate package structure for Python packages and a simple distribution process. Despite its utility, Poetry lacks public-facing files essential for sharing code, such as documentation and license files, and doesn't provide step by step instructions for setting up powerful continuous integration workflows. Another package, {\sc cookiecutter-cms}~\cite{nadenCookiecutterComputationalMolecular2024a} assists in creating Python projects, but it is much lighter weight than \skpkg. For example, it lacks an internal solution for distributing the package online. Additionally, cookiecutter-cms lacks pre-configured documents such as a README file, which acts as the front page of the software, and it does not implement automated code formatting.
For convenience,~\tabl{package-comparison} contains a summary of features of these packages compared to \skpkg.
\begin{table}
\caption{Comparison of three Python packages offering Python project scaffolding.}
\label{tab:package-comparison}
\centering
\begin{tabular}{|p{3cm}|p{3cm}|p{3cm}|p{3cm}|}
\hline
\textbf{Feature} & {\sc Poetry} & {\sc cookiecutter-cms} & {\sc scikit-package} \\
\hline
Starting code & Minimal packaging structure, no public facing files & Static public facing files, lacks detailed README & Pre-configured dynamic files with detailed README \\
\hline
Release process & Direct PyPI upload & N/A & Automated GitHub/PyPI release; conda-forge release checklist \\
\hline
Documentation & No public docs & Renders docs locally & Hosts docs with a public URL on public release \\
\hline
Testing scripts & None provided & One per project directory & Centralized scripts to manage many projects at once \\
\hline
Dependency & Advanced dependency management & No clear separation of dependencies & Separate installation requirements for docs, source code, and tests \\
\hline
Namespace support& N/A & N/A & Support nested folder structure and import package e.g. diffpy.pdffit2 \\
\hline
\end{tabular}
\end{table}

We note that our goal is not to replace any specific development tools, such as those listed in \tabl{package-comparison}.
Rather, our aim is to provide pedagogical and practical tools for scientists who, while not professionally trained software engineers, are technically minded and interested in writing more reusable and maintainable software code.
For users already familiar with Python packaging, there is an advanced Python package template developed by the Scientific Python project~\cite{scipycookie}, which offers a starting point for more advanced features, including support for selecting different build backends, often used for packages that incorporate programming languages beyond Python.
Once users become comfortable with practical tools for distribution, testing, and maintenance through \skpkg, they are encouraged to explore and adopt advanced build-backend features and tools such as \skbuild~\cite{jean_christophe_fillion_robin_2018_2565368, schreiner_2022_6946769} within the project configuration of \skpkg.
Users can also adopt the standards developed by the SPEC (Scientific Python Ecosystem Coordination) documents~\cite{scipyspec}, maintained by the Scientific Python community, to further enhance their development workflows.

The goal of \skpkg is to offer students and scientists an easy path to share code at various levels of complexity.  At the lowest level it has pedagogical examples for simple sharing such as reusing functions across files, including community recommendations and best practices,
building up in complexity all the way to sharing with the wider scientific community as a fully open-source, maintained and tested package. When the software is ready for public distribution, \skpkg provides pre-configured documents and a straightforward release process that, once it is set up, significantly reduces the time required to release and share code.
It also standardizes procedures across projects, greatly facilitating maintenance and release of multiple open source projects.

By following the suggested workflows and using the automated infrastructure provided by \skpkg, scientists can increase the impact of their research by distributing reproducible, high-quality code. 
As a result, the published code becomes significantly easier to maintain and extend, with new features contributed either internally or by the broader user community.
\skpkg can be used for new projects but also has tools for bringing existing code up to the code quality standards of \skpkg.

The development of \skpkg was motivated by the need in the Billinge research group to maintain dozens of software packages by students and short-term staff.
This requires us to maintain a level of consistency across packages and maintain uniform standards for syntax, documentation and testing.
Even updating and releasing all packages for new versions of Python was too difficult without greater uniformity across the stack. 

The starting point was the Scientific Software Cookiecutter \cite{NSLSIIScientificpythoncookiecutter2025} developed by the scientific software group at the National Synchrotron Light Source II at Brookhaven National Laboratory. 
We added continuous integration and release scripts taking advantage of GitHub workflows. 
Care was taken to generalize these and make them reusable across packages, maintaining most of the functionality in a centralized GitHub repository so updates can be easily rolled out across projects.
As a result, manual reconfiguration for each package is minimized.

Another source of increased maintenance effort was errors that were found only after release of a package. 
To minimize these we developed checklists that allow the work to be distributed across group members with minimal fluctuations in the quality of the releases. 
Where possible we incorporated automated syntax checking (linting) and uniform processes for building and deploying documentation online, including automation of application programming interface (API) documentation.

All these developments have been incorporated into \skpkg to help share these programming standards with the scientific community.
For individuals, but especially for research groups, \skpkg reduces the infrastructural overheads for open source software development and maintenance.

Using \skpkg has an additional value that it helps students and staff quickly learn Python community best practices, such as PEPs (Python Enhancement Proposals), which provide guidelines on topics such as naming conventions, line lengths, and even the use of single or double quotes.  \skpkg allows users to decide how much linting is adopted, and this can vary by the \skpkg level, with less stringent requirements at the lower levels of adoption.

Lastly, another reason we developed \skpkg was to facilitate the creation of ``branded" Python projects that support importing packages under an organization’s namespace. E.g., \utils \cite{diffpyutils}, which is imported as \texttt{diffpy.utils}, and \structure~\cite{diffpystructure}, which is imported as \texttt{diffpy.structure}~\cite{juhasComplexModelingStrategy2015a}.
This uniform naming allows us to associate sub-packages with the larger `diffpy' project, whilst conveniently maintaining the code in separate repositories. 
Since no existing tools automated this kind of namespace-based project structure, we developed \skpkg to fill that gap.
This capability can be useful for other research groups that maintain multiple packages.

\section{Overview of \skpkg}
\label{sec:overview}

\begin{figure}[1]
    \centering
    \includegraphics[width=0.9\textwidth]{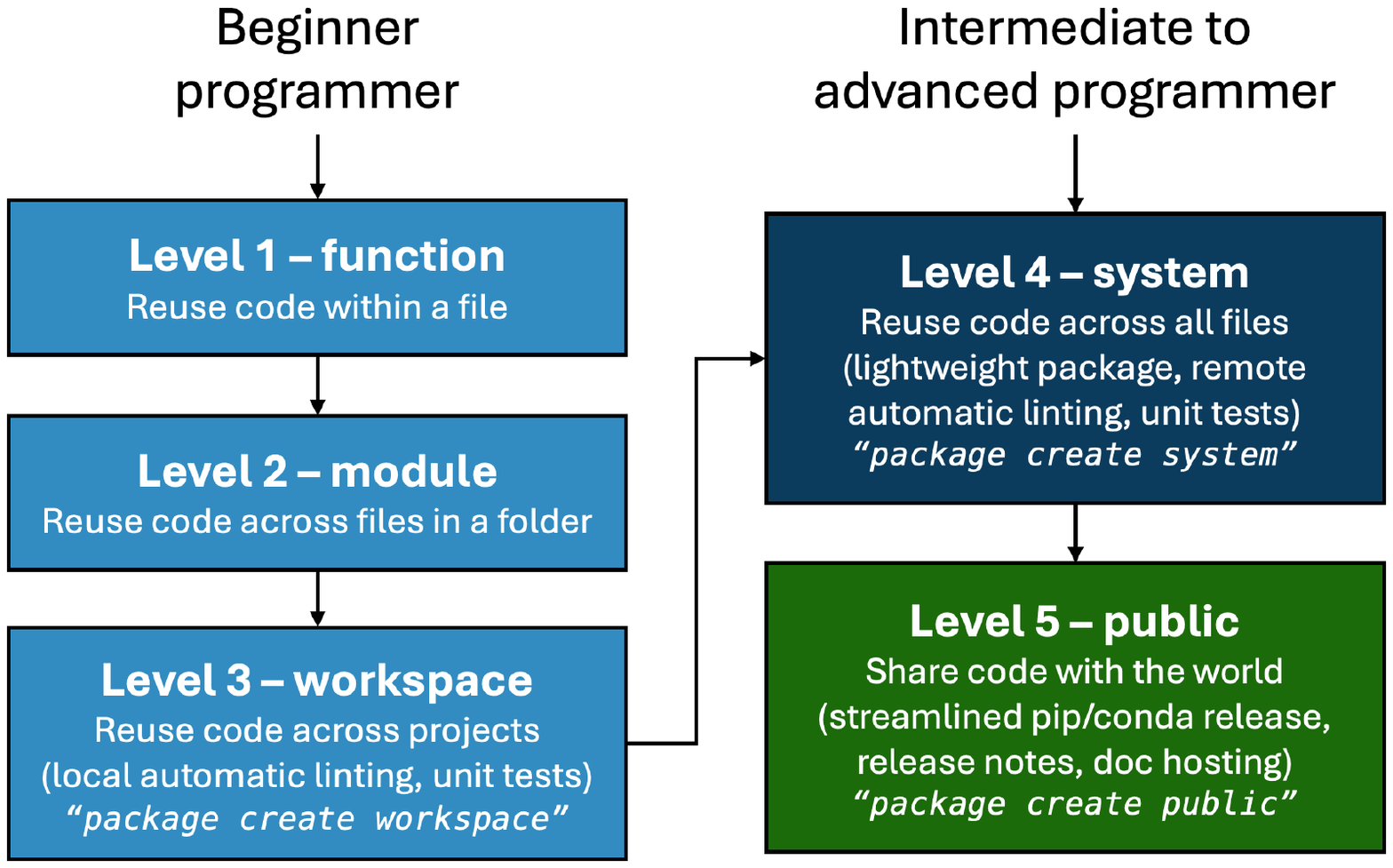}
    \caption{Diagram of 5 levels of sharing code with key features and \skpkg commands.}
    \label{fig:scikit-features}
\end{figure}

\skpkg offers step-by-step instructions for reusing and sharing code, starting from something as simple as defining and using functions, all the way to maintaining and releasing a fully documented open-source package on Python Package Index (PyPI) ~\cite{pypi} and conda~\cite{conda-docs}.
The steps are divided into five levels of shareability and complexity, allowing users to choose the level that best suits their current needs.
These are summarized in~\fig{scikit-features} and~\tabl{5-levels-of-code-share}.
\begin{table}
\caption{5 levels of code sharing.}
\label{tab:5-levels-of-code-share}
\begin{tabular}{|c|l|l|m{3cm}|}
\hline
\textbf{Level} & \textbf{Name} & \textbf{Scope} & \textbf{How to setup} \\
\hline
1 & \texttt{function} & Reuse code within the same file. & Tutorial provided \\
\hline
2 & \texttt{module} & Reuse code across files in the same directory. & Tutorial provided \\
\hline
3 & \texttt{workspace} & Reuse code across project folders. & \texttt{package create workspace} \\
\hline
4 & \texttt{system} & Reuse code across any files on the same computer. & \texttt{package create system} \\
\hline
5 & \texttt{public} & Share code as a publicly installable package. & \texttt{package create public} \\
\hline
\end{tabular}
\end{table}

\level{1}, \function, the easiest and already widely known and used, consists simply of defining functions within the same file or module.
Whilst fairly trivial, it introduces examples of community best practices. \level{2}, \module, expands on this by reusing functions across separate module files within the same directory and introduces basic importing.
\level{3}, \workspace, restructures the organization so that a block of code can be reused across multiple projects. 
This level of code reuse is already challenging for many people, who often resort to copying and pasting files.
Levels 4 and 5 involve reusing the code as an installable Python package.
\level{4}, \system, enables users to create a lightweight package so that the code can be reused across all folders on their computer and easily shared with collaborators.
\level{5}, \public, is the final step, where the source code is uploaded online so that anyone in the world can install the package, sourced from PyPI or conda-forge.

At each level the reusability increases but so does the complexity of the solution.
Users can select the level of complexity and shareability suitable to their needs.
We note that projects created at lower levels can be migrated straightforwardly to higher levels as the need arises.

\skpkg offers automated code formatting and warnings to enforce widely adopted practices outlined in PEP~8 (style guide) and PEP~257 (docstrings) to increase the quality of the code that you write.

In higher levels we utilize a library called \precommit, which integrates with both Git and GitHub.
When a new commit is created, a series of checks, known as hooks, are automatically executed.
Each hook verifies whether the files follow the defined standards and attempts to format them accordingly.
A new Git commit is created only after all hooks have passed.
Hooks that fail may involve issues that cannot be fixed automatically, such as spelling mistakes.
\precommit helps you identify the source of the problem by providing the exact files and line numbers where issues occur.
Once all hooks pass, the code can be committed to the local Git repository and then pushed to the remote GitHub repository.
In Levels~4 and~5, we also implement \precommit as part of the continuous integration in the remote GitHub repository.
This ensures that, before any new code is merged into the remote repository, \precommit runs automatically to ensure that the incoming code in the GitHub pull request meets the requisite quality standards.  
Configuration files can be edited to customize the standards applied to the current project.
\skpkg provides reasonable defaults for these configuration parameters.
Following such standards for syntax and documentation ensure that your code is more readable, maintainable and reusable by others.

\skpkg also simplifies the creation of customizable documents commonly found in a Python software package.
These documents include a README, a license, a contributor list, a code of conduct, and release notes.
For example, \texttt{README.rst} provides lightweight information telling users what the software is for as well as basic information about supported programming languages, test status, installation instructions, and guidelines for support and contributions.
If the project is contained in a repository on GitHub, the README page gets automatically displayed on the repository landing page.

When a new project is initialized with \skpkg, it prompts the user to answer a few questions, including the project maintainer's name, email address, and project description.
This metadata is then automatically populated into placeholders throughout the package’s documents.
In addition, \skpkg offers example templates for other documentation, greatly lowering the barrier to writing comprehensive but highly useful package documentation.
This allows developers to focus on writing scientific code and user guides, such as tutorials and getting-started pages.

At the highest level, Level~5, \skpkg offers an automated and streamlined process for public releases of packages by integrating with GitHub’s Continuous Integration (CI) infrastructure, GitHub Actions.

Releasing a package online can be labor intensive and time-consuming, involving manual tasks such as writing release notes, updating online documentation, and uploading the package to PyPI and conda-forge.
Using \skpkg, the project maintainer can initiate a release by creating a release tag using Git and pushing it to the GitHub repository using a local command-line tool.
Detailed instructions are in the \skpkg documentation.

This triggers a series of automated GitHub workflows that: 
\begin{enumerate}
    \item publish the release on GitHub, generating release notes in a CHANGELOG file.
    \item deploy updated documentation online with the latest tag version.
    \item uploads the package to PyPI.
\end{enumerate}
Additional steps to make the package available on conda-forge for installation with conda are also made straightforward.

It is recommended to pre-release a ``release candidate" to test all aspects of the release before making this public.
Instructions for using \skpkg to automate this task can be found in the \skpkg documentation.
As a result, scientists can save time and energy in maintaining and releasing multiple projects.

A research group often maintains multiple software packages under a single GitHub organization.
For the purposes of branding and also differentiating packages with similar names, it can sometimes be beneficial for the organization name (or some other branding name) to appear in the package name itself.
For example, the \texttt{diffpy} project lives under the 
\texttt{diffpy} org at GitHub (\url{https://www.diffpy.org/}) and contains multiple packages. 
By design, \texttt{diffpy} is prepended to all the packages, for example, \texttt{diffpy.pdfgui}, \texttt{diffpy.snmf}, and \texttt{diffpy.utils}.
An advantage of this pattern is that packages are instantly recognizable as coming from the \texttt{diffpy} project, but also that common names, such as \texttt{utils} can be overloaded and appear as \texttt{diffpy.utils}. 

Although useful, this adds complexity to the design and layout of the package structure.
\skpkg optionally supports this capability in~\level{5} if desired.
To do this, the the user enters the project name as \texttt{<namespace>.<package-name>} when using the \texttt{package create public} command.
A new package will then be created and can be imported in Python as \texttt{import <namespace>.<package-name>}. For example: \texttt{import diffpy.utils}.

Finally, we note that scientific code is often developed by teams of internal members and external collaborators.
It is in our best interest to follow best practices that increase productivity and minimize technical debt for future members.
The \skpkg public documentation (\url{https://scikit-package.github.io/scikit-package/}) includes a chapter covering best practices based on the Billinge research group's experience in developing scientific software, as well as recommendations from the Python community.
The documentation covers, among other topics, how to write effective Git commit messages and news entries used for compiling the changelog, as well as a recommended workflow for developing and requesting new features within GitHub's ecosystem.
These guides also serve as training resources for new team members, providing instructions on writing unit tests and docstrings, and on designing more useful error messages and unit tests.

\section{Getting Started with \skpkg}

Here, we briefly describe how you can use \skpkg to help with each of the five different levels of code sharing. 
The examples use the Python programming language, but the general principles can be extended to other programming languages.

\subsection{Level~1, \function: Copy-pasting code into functions}
\label{sec:level1}

Whenever you find yourself copy-pasting code in your work you should consider encapsulating that code in a function and calling the function each time it is used.
Copying and pasting code makes it difficult to maintain your software. For example, if you make an edit in one block of code you have to remember to make the same edit in all the other places where the code has been copied. 
This also can lead to errors in your data analysis, so you are strongly encouraged to develop the reflex that whenever you want to copy-paste something, to copy-paste it into a function!

If the code is reused multiple times but only within the same file, the function may be written within that same file, conventionally somewhere near the top of the document. 
\skpkg is not needed for this solution. 
However, to help novice programmers we supply an example of how to do this in the \skpkg documentation, including best-practices for naming functions and describing behavior using a docstring.  
For maximum reusability, design functions to be as modular as possible, ideally each function should only do one thing.

\subsection{Level~2, \module: Reusing code across separate modules/files}
\label{sec:level2}

Now you find your function to be so useful that you want to use it in other files (often called modules in Python) within your project. 
While it might be tempting to copy and paste the function into each module, doing so creates multiple versions of the same code that we want to avoid as discussed above. 
This copy-and-paste can be avoided by importing the function from its original module into other modules where it will be reused through importing, for example, \texttt{from <module-name> import <function-name>}.
As long as the modules are present in the same directory, this will work out of the box.
By default, Python looks for modules present in the same directory for import.
Reusing functions across directories rapidly gets more complicated and is handled in higher levels of \skpkg.

We exemplify this with a specific example where a function, \texttt{dot\_product()} is defined in a module called \texttt{shared\_functions.py} and then imported into a different module where it will be reused.
It is possible to import and use the function directly,
\begin{minted}{python}
from shared_functions import dot_product 

a = [1, 2, 3]
b = [1, 2, 3]
result = dot_product(a, b)
print(result)
\end{minted}
or you can import the entire module, in which case functions must be accessed using both the module name and function name:
\begin{minted}{python}
import shared_functions

a = [1, 2, 3]
b = [1, 2, 3]
result = shared_functions.dot_product(a,b)
print(result)
\end{minted}
In this example, Python community best practices have been followed of naming functions and modules using purely lower-case letters and ``snake\_case" where words are separated in names by underscores.

This level is described in the \skpkg documentation with examples of best practices for code syntax, documentation, and testing.

\subsection{Level~3, \textnormal{\texttt{workspace}}: Reusing code across different projects}
\label{sec:level3}

At this point, you recognize that your function is so useful that you want to use it across multiple project directories. 
However, the current file structure doesn’t support this. 
To enable code reuse across multiple projects, short of making it into a fully importable package described in Levels~4 and~5, you will need to restructure your file-folders accordingly.
This is the first place we will make use of \skpkg commands.

We take this opportunity to introduce the idea of \emph{virtual environments}. 
Virtual environments allow you to create isolated environments, each with their own software package versions installed. 
This is especially helpful when working with multiple different software programs that require conflicting versions of the same packages and is overall a good practice to use.
It is recommended to always work in virtual environments.
There are a number of options in the community, such as using Python venv \cite{python-venv}.
For uniformity, in \skpkg we use the conda~\cite{conda-docs} package and environment management system.
conda is part of the Anaconda project maintained by Continuum, and installers can be found online. 
We recommend installing Miniconda \cite{miniconda} or micromamba \cite{micromamba}, which are lightweight alternatives to the full Anaconda distribution.  
Installers for these can also be found online.

As an example of the need for virtual environments, suppose you are working on two projects, project $\alpha$ and project $\beta$. 
Project $\alpha$ requires \texttt{package-a} and project $\beta$ requires \texttt{package-b}. But, \texttt{package-a} only works with Python 3.11 and requires \texttt{numpy==1.19}, while \texttt{package-b} requires Python 3.13 and a newer version of {\sc NumPy} (i.e., \texttt{numpy==2.2.2}). 
In a standard Python setup, trying to install both packages globally leads to dependency conflicts. For example, attempting to install both packages as follows,
\begin{minted}{bash}
# Install Python 3.11-compatible dependencies, including numpy 1.19
$ pip install package-a
# Upgrade numpy to 2.2.2, which breaks packageA
$ pip install package-b

ImportError: package-a requires numpy==1.19, but found numpy 2.2.2  
\end{minted}
resulted in an import error due to the incompatibility.

With virtual environments, you can create isolated environments for your projects, each with its own Python version and only the packages it depends on.
The commands to do this for \texttt{package-a},
\begin{minted}{bash}
# Create a new environment named 'alpha-env' with Python 3.11
$ conda create -n alpha-env python=3.11
# Activate the new environment
$ conda activate alpha-env
# Install package-a into alpha-env
$ conda install package-a
\end{minted}
allows you to use \texttt{package-a} after 
activating \texttt{alpha-env}, with the following command:
\begin{minted}{bash}
$ conda activate alpha-env  # Activate env for project alpha
\end{minted}
The commands to create the environment for \texttt{package-b} are similar and it can then be run by activating the \texttt{beta-env}.
Then, when you want to work on project $\beta$, you can open a new terminal and activate the beta environment, or in the same terminal you can deactivate the alpha environment and activate the beta environment:
\begin{minted}{bash}
$ conda deactivate alpha-env  # Deactivate env for project alpha
$ conda activate beta-env  # Activate env for project beta
\end{minted}

Once you have your project's conda environment properly configured and activated, we can now install and make use of \skpkg. 
Example commands are below. The `\texttt{-n}' flag is used to specify the name of the environment, which can be any name you choose.

For example, after installing and configuring conda, typing the following commands will create a virtual environment to run \skpkg:
\begin{minted}{bash}
$ conda create -n skpkg-env scikit-package # Create environment and install scikit-package
\end{minted}

Then, each time you want to use \skpkg type
\begin{minted}{bash}
$ conda activate skpkg-env # Activate environment
\end{minted}
and all the \skpkg functionality will be available to you.

As a concrete example, when you want to initiate a new project you would:
\begin{enumerate}
    \item Open a new terminal.
    \item Navigate to the folder on your system where you want the new package to be created (e.g., in bash: \texttt{cd path/to/my/projects/folder}).
    \item Activate the \texttt{skpkg-env} environment (e.g., \texttt{conda activate skpkg-env}).
    \item Run the command \texttt{package create workspace} and follow the prompts to initialize a new workspace.
\end{enumerate}

\skpkg \texttt{workspace} is designed to help you set up a directory structure for reusing code across sub-projects on your own computer.
After running \texttt{package create workspace}, \skpkg prompts you for information that you share with it. 
In the following example the scientist wants to start a data analysis project with multiple sub-projects that will likely share code.  They would type,
\begin{minted}{bash}
$ package create workspace # Create workspace with scikit-package
    [1/1] folder_name (workspace-folder): data-analysis-projects  # Enter folder name
\end{minted}
which will generate the following directory structure:
\begin{minted}{text}
data-analysis-projects/
   |-- CODE-OF-CONDUCT.rst
   |-- README.md
   |-- requirements.txt
   |-- shared_functions.py
   |-- .gitignore
   |-- .pre-commit-config.yaml
   |-- proj-one/
       |-- __init__.py
       |-- proj_one_code.py
   |-- tests/
       |-- __init__.py
       |-- test_shared_functions.py
\end{minted}
Note that files and folders starting with a dot (\texttt{.}) at the beginning of the name are treated as ``hidden files" and don't always display by default. 
To see them in a \texttt{bash} shell type \texttt{ls -a}.
If you are viewing your files using windows explorer you will also likely have to set it up to see the hidden files.
In this paper, to avoid confusion, we will always show hidden files when we reproduce file listings. 

The generated workspace follows a simple, modular layout designed to encourage hygienic code reuse and organization. 
In order to enable code reuse across sub-projects, the \texttt{PYTHONPATH} environment variable must be set every time you open a new terminal. This can be done by navigating to the top level directory, printing the path, and defining the \texttt{PYTHONPATH} variable, for example:
\begin{minted}{bash}
$ cd /path/to/data-analysis-projects # Navigate to the workspace directory

$ pwd # Print the path to the workspace directory, copy this path
/path/to/data-analysis-projects

# For bash (e.g., Linux, macOS, Git-bash on Windows):
$ export PYTHONPATH="${PYTHONPATH}:/path/to/data-analysis-projects" # Set pythonpath variable

# For cmd or powershell on Windows:
$env:PYTHONPATH = "$env:PYTHONPATH;/path/to/data-analysis-projects"
\end{minted}
If you often work in the same directory, you can add this environment variable definition to your shell startup file (e.g., \texttt{.bashrc}, \texttt{.zshrc}, or the Windows PowerShell profile) so that it is set automatically in each new session. 
For details on how to do this, please look online or see the \skpkg documentation.

The complexity is already rising to the point where we generally prefer skipping directly to~\level{4}, but \skpkg includes Level~3 for completeness.
While this step is necessary in Level~3, it is automatically handled in~\level{4}.

After setting the environment variable, modules placed in the top-level directory can be imported and reused throughout the sub-projects.
To make use of this you must place any modules intended for reuse at the top level.
By way of example, \skpkg seeds the project with a module called \texttt{shared\_functions.py} in the correct location.
Any functions you put inside this module, or any other modules at this level of the project tree, can be reused across any sub-projects lower down in the directory tree.  
For convenience \skpkg also creates one such sub-project that can be edited to your needs.
Feel free to change the name of the project and the Python module by just renaming the folder or the filename.

You can then manually add more projects using the same pattern, for example, if you have a second sub-project it might be called \texttt{proj-two} in which case you would follow the pattern of \texttt{proj-one} and the directory tree would look something like: 
\begin{minted}{text}
data-analysis-projects/
   |-- CODE-OF-CONDUCT.rst
   |-- README.md
   |-- requirements.txt
   |-- shared_functions.py
   |-- .gitignore
   |-- .pre-commit-config.yaml
   |-- proj-one/
       |-- __init__.py
       |-- proj_one_code.py
   |-- proj-two/
       |-- __init__.py
       |-- proj_two_code.py
   |-- tests/
       |-- __init__.py
       |-- test_shared_functions.py
\end{minted}
More detailed instructions for doing this can be found in the \skpkg documentation.

In general, it is strongly encouraged to write unit tests for higher quality and more maintainable code.
The full description of how to make and use unit tests is beyond the scope of this paper, but \skpkg creates the package infrastructure for writing tests.
All tests will be located in the \texttt{tests} directory, which \skpkg seeds with an example that tests the \texttt{dot\_product()} function from \texttt{shared\_functions.py} using {\sc pytest}. 
It is encouraged to write a test for each function that is reusable.  You can also write tests for code in the sub-projects and would locate them in the same tests directory. 
A common convention, respected by {\sc pytest}, is to create a test file of name \texttt{test\_<module\_name>.py} to contain tests for all the functions held in the module of name \texttt{<module\_name>.py}. 
For example, if you have a module called \texttt{my\_file.py} you would create a test file named \texttt{test\_my\_file.py}, in the \texttt{tests} folder. 
Within that test file, you would define a test function called \texttt{test\_my\_function()} to test a function named \texttt{my\_function()}.

With this folder layout created by \skpkg, any function defined in \texttt{shared\_functions.py} can be imported and reused throughout the subdirectories. 
Further details on other files generated by \skpkg, and also on how to adapt the template for your project, can be found in the \skpkg documentation.

\subsection{Level~4, \textnormal{\texttt{system}}: Reusing code across any Python files locally}
\label{sec:level4}\

In Level~3, the module of interest is available across all projects that are subdirectories. 
However, it can be quite tedious to set the \texttt{PYTHONPATH} variable each time you open a new terminal.
Additionally, you may now want to have this module available everywhere in your local
machine, not just within the subdirectories, and want a more robust way to back it up and share it.
You can achieve this by compiling your module(s) into a locally installable Python package which is \skpkg Level~4.  
In general, Level~4 is recommended over the \skpkg~\level{3} solution described above.
Level~4 is an appropriate level for sharing code as part of a scientific publication. 
It allows others to install, run and reproduce your results.
If you expect many people to want to make use of the code as part of their research, then~\level{5} is recommended.

Backup and sharing is handled using your account on GitHub that, beyond giving you the ability to roll back code to earlier versions that Git gives us, also backs up your code to the cloud and makes it very easy to share the code with others.  
Level~4 is like~\level{5} but with lower code standards that can be developed with less overhead.
It is appropriate for projects that won't be widely shared and may be open-source but won't be formally released publicly.

After installing \skpkg (instructions in~\level{3}), follow these steps to create a new package:
\begin{enumerate}
    \item Open a new terminal.
    \item Navigate to the folder where you want the new package to be created (e.g., in bash: \texttt{cd $\sim$/dev}).
    \item Activate your conda environment containing \skpkg (e.g., \texttt{conda activate skpkg-env}).
    \item Run the command \texttt{package create system} and follow the prompts to create your new package.
\end{enumerate}

Below we show an example of the workflow where a user has the GitHub username of \texttt{sirlancelotbrave} and wants to call the package \texttt{my-science-package}. They would give the following responses (a blank response means they hit enter to accept the default value shown in parentheses):
\begin{minted}{bash}
$ package create system
  # Enter package name
  [1/6] project_name (my-package): my-science-package
  # Enter GitHub username or organization name
  [2/6] github_username_or_orgname (billingegroup): sirlancelotbrave
  # Enter name of GitHub repo (typically same as package name) 
  [3/6] github_repo_name (my-science-package): 
  # Enter name of package for display on PyPI and conda-forge
  [4/6] conda_pypi_package_dist_name (my-science-package): 
  # Name of package directory (underscore for space is required)
  [5/6] package_dir_name (my_science_package): 
  # Enter contributors' full name
  [6/6] contributors (Sangjoon Lee, Simon Billinge): Sir Lancelot, King Arthur
\end{minted}
At Level~4 you may or may not want to later distribute your package on PyPI or conda-forge.  
If you do want to distribute it there, check the package name has not been taken before choosing your package name.
In~\level{5} we will discuss in greater detail strategies for branding and naming packages.

Once finished, \skpkg generates a directory with the following structure: 
\begin{minted}{text}
my-science-package/
   |-- CODE-OF-CONDUCT.rst
   |-- LICENSE.rst
   |-- README.md
   |-- pyproject.toml
   |-- .pre-commit-config.yaml
   |-- .flake8
   |-- .gitignore
   |-- .github/
        |-- ISSUE_TEMPLATE/
            |-- bug_feature.md
        |-- workflows/
            |-- tests-on-pr.yml
   |-- requirements/
       |-- conda.txt
       |-- pip.txt
       |-- tests.txt
   |-- src/
       |-- my_science_package/
            |-- __init__.py
            |-- functions.py
   |-- tests/
       |-- test_functions.py
\end{minted}

\skpkg just creates a template for your package.
You can now copy the code you want to share into the \texttt{src/my\_science\_package} directory. 
An example is shown in~\sect{example1}, and in the \skpkg documentation.

The generated package includes a few important files to help you set up, use, and share your project.
For full details, see the \skpkg documentation, but we briefly describe some aspects here.
The \texttt{LICENSE.rst} file explains how others are allowed to use and share your code.
It is essential to have this file in order to share your code with others.  
\skpkg uses by default the BSD-3 clause license which is a widely used permissive license that maximizes the ability of others to reuse your code.
You always want to have the \texttt{LICENSE.rst} file in the top level of your code project, but if you prefer a different license, replace the text with the text of the new license.

The \texttt{pyproject.toml} file contains a set of instructions that is used during installation of your package.  
There are many options for tweaking this. 
\skpkg uses reasonable default values given other choices that are used, allowing your package to be built and installed with minimal (or no) additional effort.
It includes such things as the name of your project, the version, and the required dependencies.

You should be able to immediately install the package and make it available anywhere on your computer. 
For example, following our practice of using virtual environments, and assuming that we previously created an env named \texttt{my-science} with the correct python:
\begin{minted}{bash}
# Navigate to top-level package directory
$ cd /path/to/my-science-package 
# Install your package locally
$ conda activate my-science
$ conda install --file requirements/conda.txt
$ pip install -e . --no-deps
\end{minted}
Notice the last dot that indicates the  directory (current directory in this case) where the \texttt{pyproject.toml} file can be found.
Detailed instructions for local installation can be found in the generated \texttt{README.md} file as well as in the \skpkg documentation.
It is standard practice to include basic installation instructions and a brief project description in the \texttt{README.md}, and you are encouraged to edit it accordingly to reflect these aspects of your package.

To manage changes to your code over time and collaborate effectively, it's common to use a version control system like Git.
While a full Git tutorial is beyond the scope of this paper, you can get started by initializing a Git repository in your project directory. 
This is done by typing \texttt{git init} in the top level directory of your project.

After initialization, if you choose, you can connect your repository to GitHub to back your up code in the cloud and easily share your project with others.
Details are beyond the scope of this paper but see the \skpkg documentation for more details.

\subsection{Level~5, \textnormal{\texttt{public}}: 
\label{sec:level5}
Sharing code as publicly installable software}

At Level~5, \skpkg provides the tools to create a polished, high quality, package and tools to release it to the wider world as an open-source scientific software package.

When releasing a package for public use, you will want to ensure that the software offers well-guided documentation on what your package does, how to install it, and how to use it.
Doing so lowers the barrier to entry and helps more people use your package, boosting its impact and making your work even more rewarding!

To make it easier to install and use you will want to publish your package to PyPI~\cite{pypi} and conda-forge~\cite{conda-forge}.  
Though it makes the code easier for users, this requires more up-front work from you, the developer.
\skpkg is designed to facilitate and automate this process as much as possible.
For example, Level~5 of \skpkg automates releasing and deploying documentation using GitHub workflows. 

To use \skpkg to generate a Level~5 package, run the command:
\begin{minted}{bash}
$ package create public # run Level 5 of scikit-package    
\end{minted}

As in~\level{4}, you will be prompted to enter important information relating to your package. 
Once complete, your package will be automatically generated with the necessary files.
For specific information and assistance regarding the files, file structure, and prompts, please refer to the examples in Sections~\ref{sec:example2}-\ref{sec:example4} and the \skpkg documentation.

Once the project is set up, you can begin migrating your code into the appropriate directories in the package structure.
Instructions for organizing your code and releasing it on PyPI or conda-forge are also available in the documentation. 

Whether you're building a small utility or a full scientific software package, \skpkg provides a structured pathway for organizing, documenting, and releasing your code. 
By following its levels of code sharing, you can reduce technical debt, promote best practices in scientific programming, and ensure your software is more easily maintained and more impactful. 
In the next section, we illustrate this with specific example use cases that demonstrate how \skpkg can support real-world scientific projects.

\section{Examples using \skpkg}

Here we give four representative examples to illustrate how to use \skpkg for building shareable packages at Levels~4 and~5.  
We also include an example of how to use \skpkg to migrate your existing packages to the \skpkg standards. 
Step by step instructions are in the \skpkg documentation.
In summary,
\begin{enumerate}
\item \sect{example1}: Creating your first package at~\level{4}
    \item \sect{example2}: Creating your first public package at~\level{5}
    \item \sect{example3}: Creating a package with branded namespace import at~\level{5}
    \item \sect{example4}: Migrating an existing package to~\level{5}
\end{enumerate}

\subsection{Example 1: Creating your first package at Level~4}\label{sec:example1}

In this example, we assume \texttt{Mr Neutron} previously initiated a project called \texttt{diffraction-utils} using \skpkg~\level{3} and developed a shared class called \texttt{DiffractionObject}. 
This \texttt{DiffractionObject} is used in code analyzing various diffraction data sourced from x-ray, neutron, and electron instruments.
This class is written in a module called \texttt{diffraction\_objects.py} which is reused in Python scripts in the \texttt{scattering} sub-project folder.
\texttt{Mr Neutron}'s folder structure looks like the following:
\begin{minted}{text}
somewhere/
   |-- on/
       |-- my/
           |-- computer/
               |-- diffraction-utils/
                   |-- README.md
                   |-- diffraction_objects.py
                   |-- scattering/
                       |-- __init__.py
                       |-- neutron.py
                       |-- xray.py
                       |-- electron.py
\end{minted}

Since the project is a data analysis project, \texttt{Mr Neutron} followed common practice to place this \texttt{diffraction-utils} folder near the location of the data.
Since \texttt{Mr Neutron} wants to create a package that is shared across many different projects, it is recommended to place the package in a common directory where he keeps all his reused system-wide packages.  
This could be called anything, but following standard practice he called this directory \texttt{dev} (roughly short for code-development-area) that he placed in his home directory. 
In the example, \texttt{Mr Neutron} is working on Windows but using a \texttt{bash} terminal from the ``Git for Windows"\cite{gitforwindows} software.
This is a recommended setup for Windows users.

\subsubsection{Create a new empty project with \skpkg}

Using the bash terminal, \texttt{Mr Neutron} navigates to his \texttt{dev} directory, activates the conda environment where \skpkg is installed, and creates an empty Level~4 \system project by typing these commands: 
\begin{minted}{bash}
$ cd ~/dev
$ conda activate skpkg-env
$ package create system
\end{minted}
\texttt{Mr Neutron} enters the following values to the questions \skpkg prompts (no response indicates he hit enter to accept the default value in parentheses): 
\begin{minted}{bash}
  [1/6] project_name (my-package): diffraction-utils
  [2/6] github_username_or_orgname (billingegroup): mrneutron44
  [3/6] github_repo_name (diffraction-utils):
  [4/6] conda_pypi_package_dist_name (diffraction-utils):
  [5/6] package_dir_name (diffraction_utils): 
  [6/6] contributors (Sangjoon Lee, Simon Billinge): Mr Neutron
\end{minted}
\texttt{Mr Neutron} now sees this empty package created on the hard drive:
\begin{minted}{text}
~/dev/
    |-- diffraction-utils/
        |-- .flake8
        |-- .github
            |-- ISSUE_TEMPLATE/
                |-- bug_feature.md
            |-- workflows/
                |-- tests-on-pr.yml
        |-- .gitignore
        |-- .pre-commit-config.yaml
        |-- LICENSE.rst
        |-- README.md
        |-- requirements/
            |-- conda.txt
            |-- pip.txt
            |-- tests.txt
        |-- src/
            |-- diffraction_utils/
                |-- __init__.py
                |-- functions.py
        |-- tests/
            |-- test_functions.py        
\end{minted}
The created files are described in detail in the \skpkg documentation.

\subsubsection{Copy files to the new project directory}

At this point, \skpkg simply created an empty package with files appropriately named based on \texttt{Mr Neutron}'s responses. Next \texttt{Mr Neutron}
copies his code, the \texttt{diffraction\_objects.py} file into the \texttt{diffraction\_utils} folder.
He can do this using Windows explorer, but chooses to do it in a terminal by typing the following commands: 
\begin{minted}{bash}
$ cd ~/dev/diffraction-utils/diffraction_utils
$ cp somewhere/on/my/computer/diffraction-utils/diffraction_objects.py .
\end{minted}
\texttt{Mr Neutron} made sure to type the small dot (``.") at the end of the last command.

\subsubsection{Install package and test}

There are still a few quick steps that \texttt{Mr Neutron} needs to complete before the code is available everywhere in his computer.

First, \texttt{Mr Neutron} needs to specify the package dependencies within the \texttt{conda.txt} and \texttt{pip.txt} files in the \texttt{requirements} folder.
In this example, \texttt{Mr Neutron} enters \texttt{numpy} and \texttt{matplotlib-base} in \texttt{conda.txt} and \texttt{numpy} and \texttt{matplotlib} in \texttt{pip.txt},
one per line. After he has done this the contents of those files can be seen using the \texttt{bash} command \texttt{less}, which prints the contents of a file, 
as \texttt{Mr Neutron} does below:
\begin{minted}{bash}
$ cd requirements
$ less conda.txt
  numpy
  matplotlib-base
$ less pip.txt
  numpy
  matplotlib
\end{minted}
In general, \texttt{conda.txt} and \texttt{pip.txt} will contain the same list of dependencies.
They are the dependencies that will be installed when installing from conda and PyPI, respectively.
The reason we need separate files is that some packages have a different name on conda and PyPI, respectively.  For example, to install the lightest-weight vesrsion of \texttt{matplotlib}, for historical reasons, the conda package is called \texttt{matplotlib-base} while it is \texttt{matplotlib} on PyPI.
Any other such differences in package names across conda-forge and PyPI can also be handled this way.

Second, \texttt{Mr Neutron} must build a virtual environment and install his new \skpkg package in it.
\texttt{Mr Neutron} decides to create a new conda environment dedicated for his \texttt{diffraction-utils} package (he could have chosen to install the package in one of his existing environments).
He first deactivates the \texttt{skpkg-env} (\texttt{conda deactivate}) as he has finished the work of using \skpkg to create a new project.
He then creates a new conda environment called \texttt{diff-utils-env} using Python 3.13, installing the dependencies listed under \texttt{conda.txt}, and builds and installs his own \texttt{diffraction-utils} package, using these commands:
\begin{minted}{bash}
$ cd ~/dev/diffraction-utils
$ conda create -n diff-utils-env python=3.13
$ conda activate diff-utils-env
$ conda install --file requirements/conda.txt
$ pip install . --no-deps
\end{minted}
To test the installed package, \texttt{Mr Neutron} imports the \texttt{DiffractionObject} class from the \texttt{diffraction-utils} package in a Python module called \texttt{neutron.py}, located in the folder path \texttt{~/data-analysis/neutron-experiment/}. \texttt{Mr Neutron} writes the following line at the top of the \texttt{neutron.py} file to import the \texttt{DiffractionObject} class,
\begin{minted}{python}
# ~/data-analysis/neutron-experiment/neutron.py
from diffraction_utils.diffraction_objects import DiffractionObject
\end{minted}
without having to change any of the other code.

With the \texttt{diff-utils-env} environment activated, \texttt{Mr Neutron} can run the code by executing the command below:
\begin{minted}{bash}
# cd ~/data-analysis/neutron-experiment
$ python neutron.py
\end{minted}

We note that, since \texttt{Mr Neutron} is the developer and wants to update the code as he is working, he can install the package in his environment in  ``editable" mode (recommended) where he replaces the command \texttt{pip install .} with the following:
\begin{minted}{bash}
$ pip install -e .
\end{minted}
As long as \texttt{Mr Neutron} has activated the \texttt{diff-utils-env} conda environment, whenever he runs code anywhere on his computer it will run the version of the code it finds on disc at run-time without him having to reinstall the package.
This is very convenient for developers, but is not the preferred installation method for users.

\subsubsection{Use Git to track changes}
\label{sec:gitcontrol}

As the next step, \texttt{Mr Neutron} wants to continuously maintain the \texttt{diffraction-utils} package.
The best way to do this is to use Git.
It is beyond the scope of this article to explain Git in detail, but the main concept is that Git maintains a database on the computer with every version of every file in the user's project, so the user never loses work and can find any earlier version.
Every time \texttt{Mr Neutron} ``commits" the edits to the Git database, it stores the edits.

To set up Git, \texttt{Mr Neutron} types the following command once to initiate the Git database for the project folder (including subdirectories):
\begin{minted}{bash}
# ~/dev/diffraction-utils
$ git init
\end{minted}
To create the first commit, \texttt{Mr Neutron} executes:
\begin{minted}{bash}
$ git add .
$ git commit -m "initial commit of the package files"
\end{minted}
The \texttt{git add .} command adds all the files in the current directory (including subdirectories) to the list of files that will be committed to the database next time the user makes a commit.
The \texttt{git commit} command actually commits those edits and changes to the database with a clear commit message describing the edits.

\subsubsection{Set up \precommit hooks to automatically check syntax}
\label{sec:precommithooks}

Once the local repository is under Git control, \texttt{Mr Neutron} wants to ensure that the code is properly formatted \textit{before} committing to the Git database.
This can be done by triggering \precommit to run each time a new commit is attempted.
This is done with a pre-commit hook.
A hook automatically runs a program, or programs, before making a Git commit, every time \texttt{Mr Neutron} runs \texttt{git commit}.
\skpkg uses the \precommit package to manage this.  
To get this set up, \texttt{Mr Neutron} (who has already installed the \texttt{pre-commit} package in his environment with \texttt{conda install pre-commit}) types this command:
\begin{minted}{bash}
$ precommit install
\end{minted}
Now, every time \texttt{Mr Neutron} runs \texttt{git commit -m "<commit message>"}, he will see in the terminal the hooks being executed, for example: 
\begin{minted}{bash}
$ git commit -m "chore: implement local precommit hooks"
black...........................................................Passed
prettier........................................................Passed
docformatter....................................................Passed
\end{minted}

\texttt{Mr Neutron} proceeds to add new features and bug-fixes to the \texttt{diffraction-utils} package and commiting the changes, but sometimes \precommit hooks fail.
If this happens, \texttt{Mr Neutron} will see that the most recent commit was not written to the Git database, for example by using the \texttt{git log} command.
\texttt{Mr Neutron} can then fix those errors manually and rerun \texttt{git commit}.

At any time, \texttt{Mr Neutron} can use the \texttt{pre-commit run --all-files} command to trigger \precommit manually while fixing those errors so that he does not have to make a commit to run the checks.
As discussed in~\sect{overview}, \precommit auto-fixes based on the configurations provided in \texttt{.pre-commit-config.yaml} in the project directory. 
The output from \precommit informs \texttt{Mr Neutron} which files, and which line in the file, caused the error,
helping him fix everything up.

\subsubsection{Use GitHub to backup code online}

Having successfully made edits to the code and commit them to his local Git database,
\texttt{Mr Neutron} now wants to back-up his work online.
\skpkg is integrated with GitHub which is a cloud-based platform for uploading and sharing Git projects.

\texttt{Mr Neutron} first creates a new repository on GitHub in his user space, \texttt{mrneutron}(he could create it in any organization that he owns) and enters \texttt{diffraction-utils} for the repository name.
He selects the option to create an \textit{empty} repository (without an autogenerated \texttt{README}, \texttt{.gitignore}, or \texttt{LICENSE} file) since these files are already created with \skpkg.
\texttt{Mr Neutron} connects the local to the remote (cloud) repository by executing the following command: 
\begin{minted}{bash}
# ~/dev/diffraction-utils
$ git remote add origin https://github.com/mrneutron/diffraction-utils.git
\end{minted}
The term \texttt{origin} is an alias (name) for the remote repository.
\texttt{Mr Neutron} then runs,
\begin{minted}{bash}
$ git push --set-upstream origin main
\end{minted}
to upload the contents of the local repository to the remote repository. 
\texttt{Mr Neutron} can view the content uploaded to the remote GitHub repository.
He can also make edits directly on the remote repository and synchronize those changes with the local repository using the \texttt{git pull origin main} command.

\subsubsection{Use GitHub to share code with colleagues}

\texttt{Mr Neutron} now wants to share the code with colleagues.
The simplest way to do this is by sharing the public URL of the GitHub repository, which colleagues can use to download the code either from the website directly or through cloning, like shown below:
\begin{minted}{bash}
$ git clone https://github.com/mrneutron/diffraction-utils.git
\end{minted}
If for some reason \texttt{Mr Neutron} created the repository as a private repository rather than a public one, \texttt{Mr Neutron} can still share it with trusted colleagues by adding their GitHub usernames in the Settings page of the GitHub repository.
Or, in the spirit of open science, he can make the repository public.

As mentioned, \skpkg already created a simple \texttt{README.md} file, which by default is displayed at the repository landing page on GitHub.  The \texttt{README.md} contains basic instructions for how the colleague can clone and install the package.
\texttt{Mr Neutron} can edit the \texttt{README.md} file to make things even clearer.

This completes the \skpkg~\level{4}  example.  
In the following example, we show you how to create and maintain professional-grade software for public distribution. 
Git and GitHub offer much more than just backing up or sharing code, as described so far.
In Example~2, we explore more advanced features such as using branches, creating pull requests, and running automated workflows.

\subsection{Example 2: Creating your first package for public distribution at Level~5}\label{sec:example2}

In this example, we demonstrate how to create a new~\level{5} \public package from scratch. 
In this example the \texttt{maintainer} of the repository, who will have merge-rights, will be \texttt{Sir Lancelot}. 
To create the structure for the full featured public package he enters the following commands:
\begin{minted}{bash}
$ cd ~/dev
$ conda activate skpkg-env
$ package create public
\end{minted}
We show the responses of \texttt{Sir Lancelot} to the \skpkg prompts below. As described in Example~1,  he followed group practice and located the empty project under \texttt{$\sim$/dev}.
The text in parentheses are the default values supplied by \skpkg. 
These can be user-configured but in the example we show the \skpkg defaults.
Where there is no response \texttt{Sir Lancelot} simply hit the ``Enter" key to accept the default value:
\begin{minted}{text}
  [1/16] maintainer_name (Simon Billinge): Sir Lancelot
  [2/16] maintainer_email (sb2896@columbia.edu): sirlancelotbrave@montypy.com
  [3/16] maintainer_github_username (sbillinge): sirlancelotbrave
  [4/16] contributors (Sangjoon Lee, Simon Billinge, Billinge Group members): Sir Lancelot, Sir Robin, King Arthur
  [5/16] license_holders (The Trustees of Columbia University in the City of New York): The Knights of the Round Table 
  [6/16] project_name (diffpy.my-project): montypy
  [7/16] github_username_or_orgname (diffpy): kot-roundtable
  [8/16] github_repo_name (montypy): 
  [9/16] conda_pypi_package_dist_name (montypy): 
  [10/16] package_dir_name (montypy): 
  [11/16] project_short_description (Python package for doing science.): A Python package for the the Knights of the Round Table.
  [12/16] project_keywords (diffraction, PDF, X-ray, neutron): knights, castle, Monty, Python
  [13/16] minimum_supported_python_version (3.11): 
  [14/16] maximum_supported_python_version (3.13): 
  [15/16] Select project_needs_c_code_compiled
    1 - No
    2 - Yes
    Choose from [1/2] (1): 
  [16/16] Select project_has_gui_tests
    1 - No
    2 - Yes
    Choose from [1/2] (1): 
\end{minted}
The questions are designed to be somewhat self-describing, but what they mean and how they are used is described in detail in the \skpkg documentation.

Given the answers to the questions in the example, \texttt{Sir Lancelot} sees this folder structure created by \skpkg:
\begin{minted}{text}
~/dev/
    |-- montypy/
        |-- .codecov.yml
        |-- .codespell/
            |-- ignore_lines.txt
            |-- ignore_words.txt
        |-- .flake8
        |-- .github/
            |-- ISSUE_TEMPLATE/
                |-- bug_feature.md
                |-- release_checklist.md
            |-- PULL_REQUEST_TEMPLATE/
                |-- pull_request_template.md
            |-- workflows/
                |-- build-wheel-release-upload.yml
                |-- check-news-item.yml
                |-- matrix-and-codecov-on-merge-to-main.yml
                |-- publish-docs-on-release.yml
                |-- tests-on-pr.yml
        |-- .gitignore
        |-- .isort.cfg
        |-- .pre-commit-config.yaml
        |-- .readthedocs.yaml
        |-- AUTHORS.rst
        |-- CHANGELOG.rst
        |-- CODE-OF-CONDUCT.rst
        |-- LICENSE.rst
        |-- MANIFEST.in
        |-- README.rst
        |-- docs
            |-- Makefile
            |-- make.bat
            |-- source/
                |-- _static/
                    |-- .placeholder
            |--  api/
                |-- montypy.example_package.rst
                |-- montypy.rst
            |--  conf.py
            |-- getting-started.rst
            |-- img/
                |-- scikit-package-logo-text.png
            |--  index.rst
            |-- license.rst
            |-- release.rst
            |-- snippets/
                |--  example-table.rst
        |-- news/
            |-- TEMPLATE.rst
        |-- pyproject.toml
        |-- requirements/
            |-- build.txt
            |-- conda.txt
            |-- pip.txt
            |-- tests.txt
            |-- docs.txt
        |-- src/
            |-- montypy/
                |-- __init__.py
                |-- functions.py
                |-- version.py
        |-- tests/
            |-- conftest.py
            |-- test_functions.py
            |-- test_version.py
\end{minted}
After setting up the repository structure, 
\texttt{Sir Lancelot} adds code to the empty package by creating files in the \texttt{../src/montypy} directory. 
He adds any unit tests in the \texttt{tests} directory.
Previously written files can also be copied over from wherever they were on his hard drive, as in Example~1.
Below we show the \texttt{src} and \texttt{tests} part of the directory tree after \texttt{Sir Lancelot} completed these steps:
\begin{minted}{text}
~/dev
    |-- montypy
        |-- ...
        |-- src
            |-- montypy
                |-- __init__.py
                |-- utils.py
                |-- grail
                    |-- __init__.py
                    |-- bridge_of_death.py
                    |-- black_knight.py
                |-- version.py
        |-- tests
            |-- conftest.py
            |-- test_utils.py
            |-- test_bridge_of_death.py
            |-- test_black_knight.py
            |-- test_version.py
\end{minted}
\texttt{Sir Lancelot} made some choices about the structure of his package by choosing the directory structure within \texttt{.../src/montypy}.
This affects what the importing syntax looks like and \texttt{Sir Lancelot} made the choices so his code will be more organized and readable.
His choices resulted in import statements exemplified below,
\begin{minted}{python}
from montypy.utils import sword
from montypy.grail.bridge_of_death import questions_three
\end{minted}
assuming functions \texttt{sword()} and \texttt{questions\_three()} are defined in the modules \texttt{utils.py} and \texttt{bridge\_of\_death.py}, respectively.

Once created, the package can be put under Git control and pushed to a repository with the name \texttt{montypy} that \texttt{Sir Lancelot} creates at GitHub, following the approach in~\sect{gitcontrol}.
In the example, \texttt{Sir Lancelot} chose to create the new GitHub repository called \texttt{montypy}, under the \texttt{kot-roundtable} GitHub organization. The step-by-step tutorial for doing this is provided in the \skpkg documentation.
In this example the code could then be found at https://github.com/kot-roundtable/montypy.

\subsubsection{Recommended GitHub workflow for larger teams}

Unlike Example~1, we here introduce a forking workflow to maintain the package.
The following workflow appears initially as somewhat complicated and unnecessarily pedantic, but for group coding, following these steps quickly pays huge dividends and is worth the extra up-front effort to set up and learn in a group setting.

In our example, let's assume that \texttt{Sir Lancelot} wants a new code feature under the \texttt{grail} sub-package.  
He first creates an issue on the GitHub repository where he clicks ``New Issue" and selects the ``Bug Report / Feature Request'' template provided by \skpkg.  
He sets the title to \texttt{feat: add bucket to utils}.

On the issue page, he describes the problem, ``made a spill, need a bucket" and proposes a solution ``implement a bucket in utils.py".
GitHub provides a number for the issue, for example \texttt{\#24}, which will be used later.

In the issue title, \texttt{Sir Lancelot} added the prefix \texttt{feat:}. 
Common prefixes like \texttt{bug:} and \texttt{doc:} are used in issue titles and commit messages to help track and organize them.

A contributor to the \texttt{kot-roundtable} org, \texttt{Sir Robin}, agreed to take the issue.
\texttt{Sir Lancelot} and \texttt{Sir Robin} can discuss how to proceed in the comments thread of the issue, which results in a consensus to proceed. The issue can then be assigned by \texttt{Sir Lancelot} to \texttt{Sir Robin} so other contributors know it is under development.

In the forking workflow, \texttt{Sir Robin} will make a linked copy of the \texttt{montypy} repo under his own GitHub user namespace.  
This is called a \texttt{Fork}. 
He does this by visiting https://github.com/kot-roundtable/montypy
and clicking the \texttt{Fork} button, which results in a new linked repository at https://github.com/sirrobinbrave/montypy.

\texttt{Sir Robin} then clones this fork to his local computer with these commands:
\begin{minted}{bash}
$ cd ~/dev
$ git clone https://github.com/sirrobinbrave/montypy.git
$ cd montypy
\end{minted}

On \texttt{Sir Robin}'s local computer he has a clone of the repository that is linked to his fork, but it doesn't automatically know that the fork is linked to a repository upstream of the fork in the \texttt{kot-roundtable} org. 
\texttt{Sir Robin} then runs this command to link his local repository to the upstream one:
\begin{minted}{bash}
$ git remote add upstream https://github.com/kot-roundtable/montypy.git
\end{minted}

\texttt{Sir Robin} can now keep his local repository synchronized with all changes that are merged into the upstream repository by typing the following commands: 
\begin{minted}{bash}
$ git checkout main
$ git pull upstream main
\end{minted}

\texttt{Sir Robin} is then ready to check out a new branch to make some edits. 
He can give it any name but chooses one that he can recognize in the future:
\begin{minted}{bash}
$ git checkout -b bucket
\end{minted}
This \texttt{bucket} branch was branched from the current, most up-to-date, version of the \texttt{main} branch from the upstream GitHub repository under \texttt{kot-roundtable}.
This maximizes the probability that his edits can be merged without conflict when they are done. 
Not remembering to create new branches from a fully synchronized upstream main is one of the most common errors we see for people new to the forking workflow.

The alias (name) for remote repositories can be anything, but by convention, and in our example, the repository called \texttt{origin} links to the remote repository under username \texttt{sirrobinbrave} and the one called \texttt{upstream} is linked to the remote repository in the \texttt{kot-roundtable} org.

We recommend a workflow where branches are very granular and only contain one, or a very few, features/fixes.  
This makes it much easier to merge branches and keep the development flowing.
To the extent possible, we also recommend making branches independent of each other by creating each branch off the fully synchronized \texttt{main} branch.
These are also common mistakes of people new to the forking workflow.

In \texttt{Sir Robin}'s branch he starts by defining a \texttt{test\_bucket()} test function in \texttt{test\_utils.py}, and defining an empty function \texttt{def bucket():} in the \texttt{utils.py} module. 
\texttt{Sir Robin} then stages and commits the changes using the commands below:
\begin{minted}{bash}
$ git add tests
$ git add src/montypy/
$ git commit -m "feat: tests and function signature for bucket()"
\end{minted}

We recommend that \texttt{contributors} share code with \texttt{maintainers} as early as possible in the process,
allowing for rapid, early feedback.
In the example, \texttt{Sir Robin} does this by  creating a ``Pull request'', or PR, on the upstream repository in order to solicit the feedback. 
To do this, \texttt{Sir Robin} must first push the new branch and its local changes to his GitHub repository:
\begin{minted}{bash}
$ git push --set-upstream origin bucket
\end{minted}
The \texttt{--set-upstream} modifier, which can be shortened to \texttt{-u}, creates a permanent link between the local checked out branch (called \texttt{bucket}) with a branch of the same name on \texttt{Sir Robin}'s fork, \texttt{sirrobinbrave/montypy}.
Then, finally, to create the PR \texttt{Sir Robin} uses his browser to visit the upstream repository (\texttt{kot-roundtable/montypy)} on \texttt{GitHub} and clicks the "Pull Request" button.
There are many other ways to create the PR including integrations in IDEs such as PyCharm~\cite{pycharm} or Visual Studio~\cite{visualstudio}, using the GitHub Desktop GUI application~\cite{githubdesktop} or the \texttt{GitHub} CLI~\cite{githubcli}.

In the body of the top-level comment box of the pull request, \texttt{Sir Robin} typed instructions that he wanted the maintainer, \texttt{Sir Lancelot}, to know to help in the review.  
He also included the text,
\texttt{closes \#24},
where 24 is the issue number of \texttt{Sir Robin}'s \texttt{bucket} issue. 
Using this syntax tells GitHub to automatically close the linked issue when the pull request is approved and merged into the \texttt{main} branch.
This is a very useful feature in GitHub.

This ``PR" requests the \texttt{maintainer} of the upstream repository to ``pull" the changes in branch \texttt{bucket} in \texttt{Sir Robin}'s fork into the  \texttt{main} branch on the upstream repository.
Before doing that, \texttt{Sir Lancelot} wants to ensure that everything is just so with the code and so provides a code review on GitHub giving feedback to \texttt{Sir Robin}. 
If there are multiple \texttt{maintainers} or \texttt{contributors}, others can also review the code and suggest improvements before the changes are merged.

Next, \texttt{Sir Robin} added a ``news" file by copying \texttt{news/TEMPLATE.rst} to \texttt{news/bucket.rst} and editing it such that under the \texttt{**Added**} section he replaced \texttt{<news-item>} with "\texttt{Add bucket() in utils.py for cleaning up spills}."  
Later, at the next release, this will appear in \texttt{CHANGELOG.rst} as follows:
\begin{minted}{text}
0.1.0
=====

**Added: **

 * Add ``bucket()`` in ``utils.py`` for cleaning up spills.
\end{minted}
News items are mostly user-facing information so at each release users can see what has been added, fixed and so on, and can be written with this in mind.

\texttt{Sir Robin} then adds and commits the news file:
\begin{minted}{bash}
$ git add news/meaning.rst
$ git commit -m "chore: news"
\end{minted}
We note that it is possible to tag a PR as being a draft to let the \texttt{maintainer} know that the PR is there for feedback but is not finished and not ready to be merged.

In our example the maintainer, \texttt{Sir Lancelot}, suggests some changes in the desired behavior of \texttt{bucket()} that will allow it to be generalized and more widely usable.  He recognizes that \texttt{bucket} could be used for fetching water as well as cleaning spills.
\texttt{Sir Lancelot} suggests that \texttt{Sir Robin} add a second test to capture this new behavior.
\texttt{Sir Robin} makes these edits and adds them to the same PR by committing the changes and pushing the updated branch to his fork:
\begin{minted}{bash}
$ git push
\end{minted}
The updates automatically appear in the PR and \texttt{Sir Lancelot} is notified by GitHub of the updates and can further comment on them. 

With the tests written and capturing the desired behavior for the function, \texttt{Sir Robin} can start coding up the \texttt{bucket()} function, running the tests by typing \texttt{pytest} as described in~\level{3}.
This ``test-forward" coding approach is called ``test driven development" and often results in better designed and executed code because more thought is given to desired behavior before any coding is done on the function itself.
It is rarely done by individual programmers and is therefore not intuitive to them, but we have found that in a group context, it is a very powerful approach.

\texttt{Sir Robin} keeps coding and running tests until the tests all pass, committing and pushing to the same PR at reasonable intervals (the commits can be more frequent than the pushes).
If there is new functionality it is generally recommended to also update user documentation on the same PR to avoid forgetting it.
Anything that comes up that it doesn't make sense to fix on this PR can be captured in a new issue.

Finally, \texttt{Sir Lancelot} and \texttt{Sir Robin} agree that everything about the edits are good, all the CI is passing at GitHub, at which point \texttt{Sir Lancelot} merges the PR into  \texttt{main} on the upstream repository.

The final step is then for \texttt{Sir Robin} to synchronize his local \texttt{main} with the updated upstream \texttt{main}. 
All subsequent branches, built off a synchronized \texttt{upstream/main}, will therefore include this new feature.
Along with \texttt{Sir Robin}'s \texttt{bucket} feature, the synchronization will also fetch any edits that were merged from other contributors.

Because his edits are merged into \texttt{main} \texttt{Sir Robin} can now delete the \texttt{bucket} branch from his local computer with \texttt{git branch -d bucket}.

\texttt{Sir Robin} followed best practice and wrote a good docstring in the function definition.  
As a result, when the \skpkg continuous integration (described below) builds the documentation, the API will be automatically documented with the new function and its docstring appearing in the online docs, another nice feature of \skpkg.

\subsubsection{Continuous Integration (CI): automated GitHub workflows}

When \texttt{Sir Robin} created the PR, several separate GitHub workflows were automatically triggered.
These workflows are controlled by workflow files, located in the \texttt{.github/workflows} directory, that were created when \texttt{Sir Lancelot} started a new~\level{5} project using \skpkg.
Here are the workflows that both \texttt{Sir Robin} and \texttt{Sir Lancelot} would see:
\begin{enumerate}
    \item The first CI workflow, \texttt{Tests on PR}, runs \texttt{pytest} on all the unit tests the user wrote in the project.

    \item The second CI workflow runs \precommit to check the code quality  similar to when \precommit is run locally (described in~\sect{overview}).  To ensure that this CI test passes, get in the habit of running \precommit locally before committing and installing \precommit as a commit hook (\sect{precommithooks}).

    \item The third CI workflow uses the \texttt{Codecov} app, which adds a comment to the PR summarizing which lines of code are not covered by unit tests.  
    This workflow checks every new line of code to see if it is covered by a test and fails if insufficient tests are provided for the new code in the PR.

    \item The fourth CI workflow checks for a \texttt{news} file in the PR and is there as a reminder for this important task.
\end{enumerate}

These workflows will run on GitHub without charge for any open-source software repository that is public.
GitHub provides some free CI credits for private repositories, and it is also possible to use CI through paid plans.

For future development through this PR workflow, \texttt{Sir Robin} always waits for all CI checks to complete, either passing (green) or failing (red). For any failing tests, \texttt{Sir Robin} makes local edits on the branch, then commits and pushes those changes.
When a PR is merged into \texttt{main}, another CI workflow is triggered to ensure that the final version of the code is tested not only on Linux but also across multiple operating systems and all Python versions specified when the package was created using \skpkg.
To modify the behavior of the CI, a \texttt{maintainer} such as \texttt{Sir Robin} can modify relevant files in the \texttt{.github/workflows} directory, as described in the \skpkg documentation.

\subsubsection{Public Package Release}

We describe here what happens when \texttt{King Arthur} the code owner, and \texttt{Sir Lancelot} the maintainer, are ready to publish the \texttt{montypy} package online and share it with the wider community.
The goal is to make the package installable via the \texttt{conda install montypy} or \texttt{pip install montypy} commands.

At this point, all pull requests and issues relevant to the release must be merged and closed. 
To facilitate this process \texttt{Sir Robin} created a new GitHub issue using the \texttt{Release} template provided by \skpkg.
This issue provides a complete checklist of tasks, including testing the code, reviewing the documentation, and closing any remaining issues or pull requests, that the developer should follow to ensure a successful release.

After the checklist items are completed by \texttt{Sir Robin}, \texttt{Sir Lancelot} proceeds with the release.
\texttt{Sir Lancelot} begins by checking out the \texttt{main} branch and pulling the latest code from \texttt{upstream/main}:
\begin{minted}{bash}
$ cd ~/dev/montypy
$ git checkout main
$ git pull upstream main  # Assume forking workflow
\end{minted}

\skpkg automates the rather complex process of running releases, attempting to minimize the overhead by using reasonable defaults (which can be modified).
The release is triggered by the \texttt{maintainer}, who must have the required privileges on the GitHub repository, by simply creating a Git tag with a name with a particular pattern.
The pattern is that the tag-name follows the the semantic versioning syntax \cite{semver}.
Semantic versioning involves three numbers separated by two periods, where the three numbers indicate MAJOR, MINOR and PATCH release numbers. 
In this example, it is an initial release so \texttt{Sir Lancelot} chooses the lowest non-patch release number, \texttt{0.1.0}.:
\begin{minted}{bash}
$ git tag 0.1.0
$ git push upstream 0.1.0
\end{minted}
The automated release, up to and including the step of submitting to PyPI, is triggered on the push.

We strongly recommend doing a less public pre-release, or release-candidate (rc) before each public release.  
The release can then be tested and any issues fixed before the community becomes aware of it.
This release-candidate is a public release in the sense that the release is deployed to GitHub and to PyPI but is tagged as a pre-release.
It can be pip installed but only by specifying the full version number and can only be found on PyPI and GitHub by some digging.

In our example, \texttt{Sir Lancelot} could make a pre-release of the \texttt{0.1.0} release by running,
\begin{minted}{bash}
$ git tag 0.1.0-rc.0
$ git push upstream 0.1.0-rc.0
\end{minted}
with exactly this format (included dashes and dots).  

The default release obtained by typing \texttt{pip install montypy} remains as the existing release, but the pre-release can be installed in a test environment by explicitly specifying the release number, \texttt{pip install montypy==0.1.0-rc.0}
If needed, a second rc release with some problems fixed would be numbered \texttt{0.1.0-rc.1}, and so on.

Whether it is a release-candidate or a full release, the GitHub tag pushed to the upstream repository  triggers a series of GitHub workflows, including a check to verify whether the user executing the tag is authorized.
When the package was created, \texttt{Sir Lancelot} entered \texttt{sirlancelotbrave} in response to the \texttt{maintainer\_github\_username} prompt. 
The \texttt{maintainer\_github\_username} specified person (this can be updated manually in the workflows) is the only GitHub user authorized to run this release workflow. 
Otherwise, the workflow will fail and the release process will not proceed.
This ensures that only an authorized person can release the code.
Once the workflow succeeds, it will then create a new pre-release/release on GitHub and publish the package to PyPI.

We recommend to also make packages available on \texttt{conda-forge} which can host not just python and has powerful methods for checking the dependency tree of all packages. For hints for how to do this please see the \skpkg documentation.

After verifying the package is available and functional, \texttt{Sir Robin} who created the GitHub Release issue, can close it which then completes the release lifecycle for the version.

\subsection{Example 3: Creating a package with namespace import at Level~5}\label{sec:example3}

In Example~2, we demonstrated how a package called \texttt{montypy}, containing the sub-package \texttt{grail}, was created and maintained by \texttt{Sir Lancelot} and \texttt{Sir Robin} in the group of \texttt{King Arthur}.
Now, hypothetically, \texttt{King Arthur} is interested in creating another sub-package called \texttt{meaning} in the \texttt{montypy} package.
To prevent a single package from becoming too bloated, King Arthur has decided that each module should be maintained in a separate repository as a separate package.
King Arthur uses the full benefit of \skpkg with namespace imports, where each package can be developed and imported as \texttt{<namespace-name>.<package-name>}. In this case, the package containing the \texttt{grail} module can be named \texttt{montypy.grail}, and the package for the \texttt{meaning} module can be named \texttt{montypy.meaning}.
\texttt{Sir Lancelot} continues to be the maintainer in the group.
In Example~3, we show how to create two separate packages, each hosted in its own GitHub repository and managed with a separate local Git database, while still sharing the common namespace \texttt{montypy}.

In a conda environment where \skpkg is installed, \texttt{Sir Lancelot} enters the following responses to the \skpkg prompts:
\begin{minted}{text}
$ package create public
  [1/16] maintainer_name (Simon Billinge): Sir Lancelot
  [2/16] maintainer_email (sb2896@columbia.edu): sirlancelotbrave@montypy.com
  [3/16] maintainer_github_username (sbillinge): sirlancelotbrave
  [4/16] contributors (Sangjoon Lee, Simon Billinge, Billinge Group members): Sir Lancelot, King Arthur
  [5/16] license_holders (The Trustees of Columbia University in the City of New York): The Knights of the Round Table 
  [6/16] project_name (diffpy.my-project): montypy.grail
  [7/16] github_username_or_orgname (diffpy): kot-roundtable
  [8/16] github_repo_name (montypy.grail): 
  [9/16] conda_pypi_package_dist_name (montypy.grail): 
  [10/16] package_dir_name (montypy.grail): 
  The other inputs are the same as those shown in Example 2
  ...
\end{minted}
In the prompts above, once \texttt{Sir Lancelot} has entered the value of \texttt{project\_name} of \texttt{montypy.grail} to get the correct namespace structure.  After that, he used the default values generated by \skpkg based on his first response by pressing the ``Enter" key for \texttt{github\_repo\_name}, \texttt{conda\_pypi\_package\_dist\_name},
and \texttt{package\_dir\_name}.

\texttt{Sir Lancelot} then runs \skpkg again with the following prompt responses:
\begin{minted}{text}
$ package create public
  [1/16] maintainer_name (Simon Billinge): Sir Lancelot
  [2/16] maintainer_email (sb2896@columbia.edu): sirlancelotbrave@montypy.com
  [3/16] maintainer_github_username (sbillinge): sirlancelotbrave
  [4/16] contributors (Sangjoon Lee, Simon Billinge, Billinge Group members): Sir Lancelot, Sir Robin, King Arthur
  [5/16] license_holders (The Trustees of Columbia University in the City of New York): The Knights of the Round Table 
  [6/16] project_name (diffpy.my-project): montypy.meaning
  [7/16] github_username_or_orgname (diffpy): kot-roundtable
  [8/16] github_repo_name (montypy.meaning): 
  [9/16] conda_pypi_package_dist_name (montypy.meaning): 
  [10/16] package_dir_name (montypy.meaning): 
  The other inputs are the same as shown in Example 2
  ...
\end{minted}

After moving over existing code, this resulted in the following folder structure for \texttt{montypy.grail},
\begin{minted}{text}
~/dev/
    montypy.grail
    |-- src
        |-- montypy
            |-- __init__.py
            |-- grail
                |-- __init__.py
                |-- bridge_of_death.py
                |-- black_knight.py
                |-- utils.py
                |-- version.py
    |-- ...
\end{minted}
and this for \texttt{montypy.meaning}:
\begin{minted}{text}
~/dev/
    montypy.meaning
    |-- src
        |-- montypy
            |-- __init__.py
            |-- meaning
                |-- __init__.py
                |-- mr_creosote.py
                |-- the_bucket.py
                |-- utils.py
                |-- test_version.py
    |-- ...
\end{minted}
These can be put under Git control and linked to GitHub repositories, worked on and ultimately released.
Now, anyone can install and use these two packages independently. 
They can be imported into a Python script as shown below:
\begin{minted}{python}
from montypy.grail import bridge_of_death
from montypy.meaning import mr_creosote

bridge_of_death.sword()
mr_creosote.better()
\end{minted}
Each project is hosted on GitHub in separate repositories at https://github.com/kot-roundtable/montypy.grail and https://github.com/kot-roundtable/montypy.meaning.
Each project also has access to the benefits available to a~\level{5} \public package shown in Example~2.

Currently, a common \texttt{utils.py} module exists in both \texttt{montypy.grail} and \texttt{montypy.meaning}.
As a useful next step, it would be beneficial to create another package called \texttt{montypy.utils} so that this module can be shared across all projects maintained by the \texttt{kot-roundtable} GitHub organization, led by \texttt{King Arthur}.
This is described below.

\subsection{Example 4: Migrating an existing package to Level~5}\label{sec:example4}

In the final example, we demonstrate how you can migrate an existing legacy Python package to the~\level{5} \public standard.
As in Example~2, we adopt the forking GitHub workflow where \texttt{Sir Lancelot} is the maintainer and \texttt{Sir Robin} is the contributor.
The migration process is divided into two parts.
The first involves fixing legacy issues that allow the existing code to pass \precommit. 
Second, a new package directory structure is created and files from the old package are migrated over into the new package directory tree.
The files fall into three categories, whether they are present just in the new package, just in the old package or in both.
\skpkg has instructions for how to use Git to help with this transition.

Here we assume the legacy project is called \texttt{flying-circus} and it exists on GitHub under the \texttt{kot-roundtable} organization.
In the past, \texttt{Sir Robin} has forked the \texttt{flying-circus} remote repository and cloned it onto his local computer in the \texttt{$\sim$/somewhere/on/my/computer} folder using the forking workflow that was described in Example~2.
As a result, \texttt{Sir Robin}'s file system has the following structure:
\begin{minted}{text}
~/somewhere/
   |-- on/
       |-- my/
           |-- computer/
               |-- flying-circus/
                   |-- .gitignore
                   |-- .git/
                   |-- README.md
                   |-- setup.py
                   |-- requirements.txt
                   |-- flying_circus/
                       |-- __init__.py
                       |-- surreal.py
                       |-- tests/
                           |-- test_surreal.py
\end{minted}
It is a legitimate package that has been released to the public, but it is not to the \skpkg~\level{5} standards.  
For example, it uses the not-recommended \texttt{setup.py} rather than the preferred \texttt{pyproject.toml} to handle the package build.

\subsubsection{Lint code with black}

The first step of bring it to standard involves autolinting the existing code with \texttt{black} to \skpkg syntax standards.
As usual, \texttt{Sir Robin} starts by making sure his local repository is properly synchronized, then creates a new branch called \texttt{black-edits} for this work:
\begin{minted}{bash}
$ cd ~/somewhere/on/my/computer/flying-circus
$ git checkout main
$ git pull upstream main
$ git checkout -b black-edits
\end{minted}
He then activates the \texttt{skpkg-env} conda environment for doing the work and installs \texttt{black}:
\begin{minted}{bash}
$ conda activate skpkg-env
$ conda install black
\end{minted}
The configuration information for the \texttt{black} tool is held in a \texttt{pyproject.toml} file at the top level of the repository.
Because Sir Robin doesn’t have a \texttt{pyproject.toml} file, he will need to create
one with the desired configuration. 
Following the \skpkg instructions, he creates a \texttt{pyproject.toml} file (using \texttt{touch pyproject.toml}) and adds this code block:
\begin{minted}{text}
[tool.black]
line-length = 79
include = '\.pyi?$'
 exclude = '''
 /(
     \.git
 | \.hg
 | \.mypy_cache
 | \.tox
 | \.venv
 | \.rst
 | \.txt
 | _build
 | buck-out
 | build
 | dist
 | blib2to3
 | tests/data
 )/
 '''    
\end{minted}
If he wants he can modify the configuration to conform to the standards of the project he is working on, but for \texttt{flying-circus} he is happy to take the \skpkg defaults.

To run the auto-linter on the code, \texttt{Sir Robin} types the command \texttt{black .}, where the dot means current directory and all subdirectories.
This makes (generally) safe, automatic updates to all the code files it finds in the project.
He can then commit the changes and make a PR so that \texttt{Sir Lancelot}, the project \texttt{maintainer}, can review and merge.
He is careful not to make any manual edits in this PR so he tells \texttt{Sir Lancelot} that all the edits are from \texttt{black}, which makes it easy for \texttt{Sir Lancelot} to merge it.

\subsubsection{Setup \precommit  to format code}

Next we continue with more linting activities beyond autolinting so that all the checks in \precommit pass.
At this point, following the \skpkg instructions, \texttt{Sir Robin} creates an empty~\level{5} \texttt{public} project in the current directory by typing the following:
\begin{minted}{bash}
$ cd ~/somewhere/on/my/computer/flying-circus  # he should already be here
$ conda activate skpkg-env
$ package create public
\end{minted}
He answers the questions as in Example~2, giving \texttt{flying-circus} as the package name.  
This results in a new subdirectory in \texttt{flying-circus} called \texttt{flying-circus} (the same name).
The steps are a bit involved and are discussed in detail in the \skpkg documentation. 
His directory structure now looks like the following:
\begin{minted}{text}
somewhere/
   |-- on/
       |-- my/
           |-- computer/
               |-- flying-circus/
                   |-- .gitignore
                   |-- .git/
                   |-- pyproject.toml
                   |-- README.md
                   |-- setup.py
                   |-- requirements.txt
                   |-- flying_circus/
                       |-- __init__.py
                       |-- surreal.py
                       |-- tests/
                           |-- test_surreal.py
                   |-- flying-circus/  # Level 5 empty folder
                       |-- .codecov.yml
                       |-- .codespell/
                           |-- ignore_lines.txt
                           |-- ignore_words.txt
                       |-- .flake8
                       |-- .github/
                       |-- .gitignore
                       |-- .isort.cfg
                       |-- .pre-commit-config.yaml
                       |-- .readthedocs.yaml
                       |-- src/
                       |-- tests/
                       |-- requirements/
                       |-- pyproject.toml
                       |-- ...
\end{minted}

To begin with we need to take the \precommit configuration files from the new package created by \skpkg and place them in the old package. 
These include the \texttt{.pre-commit-config.yaml}, \texttt{.isort.cfg}, \texttt{.flake8}, and so on.
After doing this, \texttt{Sir Robin}'s
directory structure looks like this:
\begin{minted}{text}
somewhere/
   |-- on/
       |-- my/
           |-- computer/
               |-- flying-circus/
                   |-- .codespell/
                       |-- ignore_lines.txt
                       |-- ignore_words.txt
                   |-- .flake8
                   |-- .isort.cfg
                   |-- .pre-commit-config.yaml
                   |-- .gitignore
                   |-- .git/
                   |-- pyproject.toml
                   |-- README.md
                   |-- setup.py
                   |-- requirements.txt
                   |-- flying_circus/
                       |-- __init__.py
                       |-- surreal.py
                       |-- tests/
                           |-- test_surreal.py
                   |-- flying-circus/  # Level 5 empty folder
                       |-- .codecov.yml
                       |-- .codespell/
                           |-- ignore_lines.txt
                           |-- ignore_words.txt
                       |-- .flake8
                       |-- .github/
                       |-- .gitignore
                       |-- .isort.cfg
                       |-- .pre-commit-config.yaml
                       |-- .readthedocs.yaml
                       |-- .gitignore
                       |-- src/
                       |-- tests/
                       |-- requirements/
                       |-- pyproject.toml
                       |-- ...
\end{minted}
With this done, \texttt{Sir Robin} installs \precommit in his environment and then runs it with this command:
\begin{minted}{bash}
$ pre-commit run --all-files
\end{minted}
He sees many errors raised by \precommit.
He will fix them and get the cleaned code reviewed and merged by \texttt{Sir Lancelot} on a bunch of different branches and PRs, but to avoid these changes inadvertently breaking the code at the upstream repository, \texttt{Sir Lancelot} creates a new branch on the \texttt{kot-roundtable/flying-circus} repository at GitHub, calling it \texttt{migration}.  
All PRs that \texttt{Sir Robin} creates now he will request to have them merged into the \texttt{migration} branch, which he will use the same way he was using \texttt{upstream/main}, keeping it synchronized and building new branches off the his local \texttt{migration} branch that is synchronized with \texttt{upstream/migration}.
Only at the end, when everything tested and working, will \texttt{Sir Lancelot} merge \texttt{flying-circus/migration} into \texttt{flying-circus/main}.

\texttt{Sir Robin} continues his work to fix errors raised by \precommit.
For each category of errors, \texttt{Sir Robin} creates a dedicated branch, grouping similar fixes together, with the following commands:
\begin{minted}{bash}
$ git checkout --track upstream/migration
$ git pull upstream migration
$ git checkout -b pre-commit-<theme>
\end{minted}
For example, a branch called \texttt{pre-commit-spelling} contained spelling fixes, while another branch, \texttt{pre-commit-flake8-line} contained fixes of line length errors raised by \texttt{flake8}.
These were pushed to \texttt{Sir Robin}'s fork and PRs created into \texttt{flying-circus/migration} branch for review and merge by \texttt{Sir Lancelot}, as we have described.
More granular branches make \texttt{Sir Lancelot}'s job to review and merge changes much easier.

\subsubsection{Setup local CI after migrating essential files}

With the package now passing all \precommit checks and local tests, it is time to start migrating it to the new package structure created by \skpkg.
We do this by copying files from the old package into the directory structure created by \skpkg.

The old package was under git control.  
We have found that the best way to do the migration is to first move the Git database from the existing project directory to the new~\level{5} package.
This retains the entire git history of the old project, but places it in the new package structure created by \skpkg.
After we do this, the git controlled \texttt{flying-circus} package is now the new package and the files in the old package are no longer under git control, until we move them over.

To move the Git database over, \texttt{Sir Robin} executes the below commands:
\begin{minted}{bash}
$ cd flying-circus  # Enter Level 5 directory
$ mv ../.git .      # Move Git database from old to new directory
\end{minted}
When \texttt{Sir Robin} types \texttt{git status}, he sees files listed as \texttt{deleted}, \texttt{added}, and \texttt{modified}.
This is from the point of view of the Git database rather than actual reality.
\begin{enumerate}
    \item \texttt{deleted}: These are files that exist in the Git database but are no longer present in the new package structure (e.g. project source code).
    
    \item \texttt{Untracked files}: These are files that Git finds in the new package structure that do not yet exist in the Git database (i.e. new files introduced by \skpkg).
    
    \item \texttt{modified}: These files exist in both the Git database and the new package structure, but their contents differ.
    
    \item (not listed): Files that exist in both locations and are identical will not appear in the \texttt{git status} output.
\end{enumerate}

As an example, the code in the old package (\texttt{surreal.py} and \texttt{test\_surreal.py}) hasn't been moved over so will show in the list as \texttt{deleted}.
The \texttt{requirements} and \texttt{src} directory trees don't exist in the old package and will be listed as \texttt{untracked}.
\texttt{pyproject.toml} is in both places but with different content, so will show as \texttt{modified}.
And, assuming that \texttt{Sir Robin} didn't modify any of the \texttt{flake8} defaults, the \texttt{.flake8} will not appear in the list at all as, from the point of view of the git database, it has not changed.

\texttt{Sir Robin} can then start the work of removing all the issues from the \texttt{git status} list bit by bit. 
He first copies over the code files with:
\begin{minted}{bash}
$ cp -n ../flying_circus/surreal.py ./src/flying_circus/ 
$ cp -n ../flying_circus/tests/test_surreal.py ./tests
\end{minted}

The \texttt{-n} modifier in the \texttt{cp} command stands for ``no clobber" and ensures that, if there is a file in the destination of the same name, the \texttt{cp} command will fail and the destination file will not be overwritten.  Clearly this is not needed here, but it does no harm, and is a good habit to avoid errors in this migration process.
When copying over entire directories, you would use the command \texttt{cp -n -r} where the \texttt{-r} means recursively and the copy command will copy all the subdirectories and their contents. 
Here the \texttt{-n} can be very important.  
However, pay attention to outputs from the \texttt{cp} command and make notes of any clashes that need to be manually resolved.

After the code is moved over, it should be possible to build the code in the new package and have the tests pass.
\texttt{Sir Robin} updates the files in the \texttt{requirements} directory, \texttt{docs.txt}, \texttt{conda.txt}, \texttt{pip.txt}, and \texttt{tests.txt}, adding any dependencies that are needed for the code to run.
\texttt{Sir Robin} can then confirm everything works with the code in the new package by creating a new environment and running the tests with the following commands:
\begin{minted}{bash}
$ conda create -n flying-circus-env python=3.13
$ conda install --file requirements/conda.txt
$ conda install --file requirements/tests.txt
$ pip install -e . --no-deps
$ pytest
\end{minted}
These changes can be committed, pushed and turned into a PR into the \texttt{upstream/migration} branch.

Now, the tests are passing locally, but for them to pass in the CI on GitHub some more files need to be added to the Git database.  
In particular, add and commit the \texttt{.github/} directory, as well as the \texttt{src/}, \texttt{tests/}, and \texttt{requirements/}, directories.
If these are added to the PR, the unit tests should now also pass in the CI.

After the GitHub automated workflows pass, \texttt{Sir Lancelot} can review and merge the \texttt{sirrobinbrave/setup-CI} branch to the \texttt{upstream/migration} branch.

\texttt{Sir Robin} can now move handwritten documentation files, such as tutorials under the \texttt{doc} directory and the \texttt{README}.
First, \texttt{Sir Robin} synchronizes his local \texttt{migration} branch and creates a new branch called \texttt{doc}.

Files that show as \texttt{updated} need to be handled carefully.
They exist in the old package and the new package but with different contents and need to be manually merged by \texttt{Sir Robin}.

Some other files also need careful merging when they contain similar content but have a different name between the packages.  
For example, in this case the \texttt{README.md} file in the old package is renamed to \texttt{README.rst}
in the new one.  \texttt{README.md} therefore shows up as \texttt{deleted} and \texttt{README.rst} as \texttt{untracked} in the \texttt{git status} list.  
In this case \texttt{Sir Robin} will add and commit \texttt{README.rst} but then open them both in a text editor and copy any text over from the old \texttt{README.md} to the new \texttt{README.rst}, before finally removing the \texttt{README.md} from the Git database using \texttt{git rm README.md}.
In this example, \texttt{setup.py} receives a similar treatment as its functionality is replaced by \texttt{pyproject.toml} in the new project, though some of the information in the old \texttt{setup.py} may still be needed in the new \texttt{pyproject.toml} and so is manually merged in an editor/IDE by \texttt{Sir Robin}.

The work is considered finished when:
\begin{enumerate}
    \item All files showing as \texttt{deleted} that need to be preserved have been moved from the old to the new structure directory structure.
    \item All files showing as \texttt{deleted} that are unwanted in the new package have been removed from the Git database using \texttt{git rm <filename>}.
    \item All \texttt{untracked} files created by \skpkg have been \texttt{git add}ed and \texttt{git commit}ted.
    \item All \texttt{modified} files that exist in both the old and new locations have been reviewed and the contents merged.
    \item All resulting pull requests have been reviewed and merged by \texttt{Sir Lancelot}.
\end{enumerate}

Finally, assuming all tests are passing and he is happy, \texttt{Sir Lancelot}, can merge the \texttt{upstream/migration} branch into the \texttt{upstream/main} default branch.

In his computer, \texttt{Sir Robin} can then clean and organize things.  
He updates his local \texttt{main} from \texttt{upstream/main} and moves the new Level~5 package directory to his global \texttt{dev} folder:
\begin{minted}{bash}
# Move Level 5 directory to ~/dev
$ mv somewhere/on/my/computer/flying-circus/flying-circus ~/dev  
\end{minted}
As a result \texttt{Sir Robin} sees that his package has been moved to the appropriate place with all his code, resulting in the directory structure,
\begin{minted}{text}
~dev/
   |-- flying-circus/
       |-- .codecov.yml
       |-- .codespell/
       |-- .flake8
       |-- .github/
       |-- .gitignore
       |-- .isort.cfg
       |-- .pre-commit-config.yaml
       |-- .readthedocs.yaml
       |-- AUTHORS.rst
       |-- CHANGELOG.rst
       |-- CODE-OF-CONDUCT.rst
       |-- LICENSE.rst
       |-- MANIFEST.in
       |-- README.rst
       |-- docs/
       |-- news/
       |-- pyproject.toml
       |-- requirements/
       |-- src/
          |-- flying_circus/
               |-- __init__.py
               |-- surreal.py
       |-- tests/
           |-- test_surreal.py
\end{minted}
and he is ready to continue to maintain and develop \texttt{flying-circus} as a community open source project.

\section{Conclusion}

\skpkg is designed to help researchers reuse and share scientific code by providing a roadmap.
This roadmap defines five levels of code sharing: \function, \module, \workspace, \system, and \public, ranging from reusing individual functions within a file to releasing installable, professional grade software packages.
To support onboarding, we provide comprehensive tutorials and migration examples which also help researchers adopt best community practices for software development.
When a package is ready for public release, \skpkg reduces maintenance overhead by streamlining the release process through automation that manages publishing, documentation updating, and version control.
It also supports namespace package imports, which allow organizations to maintain consistent branding and allow them to use package names that would otherwise conflict with existing packages.
By lowering the effort required for maintenance and promoting code reuse, \skpkg helps improve reproducibility and amplify the impact of scientific research.

\ack{Acknowledgments}
We thank the NSLS-II scientific-python-cookiecutter project for providing the starting point for our development. We are also grateful to the conda-forge volunteers for their support of the open source community.
We thank the cookiecutter project for providing a framework to scaffold new projects. 
We thank All the Billinge-group contributors to our code base, but especially Pavol Juhas and Chris Farrow for their earlier development of software in the Billinge group, which provided the basis for implementing and testing \skpkg, including the \texttt{diffpy.} namespace structure.
We also thank Yuchen Xiao for the development of user-configurable defaults, Yucong Chen, Tina Na Narong, and Sam Andrello for their constructive feedback, which helped improve \skpkg and its documentation.
Work in the Billinge Group was supported by the U.S. Department of Energy, Office of Science, Office of Basic Energy Sciences (DOE-BES) under contract No. DE-SC0024141.

\section{Contributions}
S.J.L.B. proposed the research topic.
S.L. implemented the five levels of code sharing, wrote the documentation, and designed the figures and tables.
S.J.L.B. reviewed and merged pull requests on GitHub, reviewed all parts of the manuscript, and co-wrote the examples with S.L..
C.M. led the ``Getting Started" section of the manuscript, provided feedback on the levels of sharing and example, and created the logo.
S.L., S.J.L.B., C.M. wrote the manuscript and all authors reviewed it.
A.Y. led the development of the code migration framework and wrote an internal guide for migrating a legacy package to the public package standard. T.Z. and S.L. wrote the GitHub release workflows. All authors contributed to code review.

\bibliographystyle{iucr}
\bibliography{bg-pdf-standards,billinge-group-bib,sbl-scikit-package,hand-entered}

\newpage

\end{document}